\documentclass[a4paper,11pt]{article}
\usepackage{moriond,epsfig,amsmath,amsfonts,amssymb}

\def\Journal#1#2#3#4{{#1} {\bf #2}, #3 (#4)}

\def\PLB{{\em Phys. Lett.}  B}
\def\PRL{\em Phys. Rev. Lett.}
\def\PRD{{\em Phys. Rev.} D}
\def\ZPC{{\em Z. Phys.} C}
\def\NCA{{\em Nuovo Cim.} A}

\def\ra{\rightarrow}

\begin{document}
\vspace*{4cm}

\title{STRONG ANOMALIES IN KAON DECAYS}

\author{S. TRINE}

\address{Institut de Physique Th\'eorique et Math\'ematique,\\
Universit\'e Catholique de Louvain, Chemin du Cyclotron, 2,\\
B-1348 Louvain-la-Neuve, Belgium}

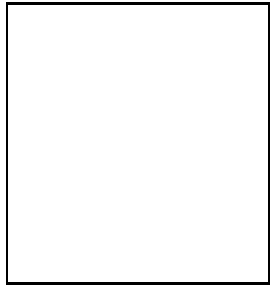
\maketitle

\abstracts{We point to a potentially large contribution of the QCD trace anomaly to the isoscalar pion pair production in Kaon decays. This contribution arises from the long-distance evolution of a 7-dimensional quark-gluon operator generated by short-distance QCD corrections to the $\Delta S =1$ Fermi Hamiltonian.}

\section{Introduction}

In addition to the well-known anomaly in the global $U(1)$ axial symmetry, massless QCD also exhibits a breaking of the invariance under dilatations at the quantum level. This is known as the trace anomaly. While the axial anomaly is related to a pseudoscalar combination of two gluon fields, the trace anomaly is associated with a scalar combination of two such fields.

Their implications in the hadronic world are numerous. For example, in the quarkonium transitions $\Psi ' \ra J/\Psi \eta$ and $\Psi ' \ra J/\Psi \pi \pi$, it has been shown\,\cite{Voloshin80} that the gluon conversions
into $\eta$ and $\pi \pi$ are governed by the axial and trace anomalies respectively.

Similarly, the possibility of trace anomaly dominance has been investigated in weak K decays.\,\cite{Gerard01} In this talk, we show how an enhancement of the isoscalar pion pair production due to trace anomaly effects may be induced by strong interaction corrections to the $\Delta S =1$ Fermi Hamiltonian.

\section{A trace anomaly contribution to the $\Delta I=1/2$ rule in Kaon decays}

The lowest order $\Delta S=1$ quark-gluon $(qG)$ effective Hamiltonian that contains the trace anomaly $(TA)$ operator reads:
\begin{equation}
H_{TA}^{(qG)} \thicksim \bar{s}_{R} d_{L} \, G_{\mu \nu}^{a} G^{\mu \nu , a},
\label{eq:QGTA}
\end{equation}
where $G_{\mu \nu}^{a}$ denotes the gluon field strenght tensor and the subscript $L(R)$ refers to left(right)-handed spinors.

This 7-dimensional operator is generated by leading order short-distance (SD) QCD corrections to the $\Delta S =1$ Fermi Hamiltonian, at next-to-leading order in the momentum expansion.\,\cite{Penin93}
Yet the bulk of its contribution is expected to come from long-distance (LD) physics, as the anomaly is a non-perturbative phenomenon. Its effects on the $\Delta I=1/2$ amplitudes could thus be larger than first expected from power counting arguments.

The LD evolution is computed in a truncated non-linear sigma model approach. The initial operator (\ref{eq:QGTA}) is bosonized into $U^{ds}$ at the confining scale. The unitary matrix $U$ is related to the octet of light pseudoscalar fields $\pi^{a}$ through
$U=\exp (i \sqrt{2} \Pi / F_{\pi}) \, , \ \Pi=\sum_{a=1}^{8} \lambda^{a} \pi^{a} / \sqrt{2} \, .$
$F_{\pi}$ is the neutral pion decay constant and $\lambda^{a}$ are the Gell-Mann matrices with normalization
$tr \lambda^{a} \lambda^{b} =2 \delta^{ab}$. Strong interaction corrections are modelled by the usual $O(p^{2})$ non-linear Lagrangian
$\mathcal{L}_{NL} = (F_{\pi}^{2}/4) tr(\partial _{\mu}U \partial ^{\mu} U^{\dagger})
                   + r(F_{\pi}^{2}/4) tr(MU^{\dagger}+UM)$,
where $M$ is the real, diagonal, light quark mass matrix.

Using the background field method with a cut-off regularization, we obtain the following result at the one-loop level:\,
\footnote{Note that a similar computation has already been performed in another context.\,\cite{Hambye98}}
\begin{eqnarray}
U^{ds} & \ra & (1+a) \, U^{ds}
\nonumber \\
& + & (b/2F_{\pi}^{2}) \, U^{ds} \, tr(\partial _{\mu}U \partial ^{\mu} U^{\dagger}+r(MU^{\dagger}+UM)) \nonumber \\
& + & (3b/2F_{\pi}^{2}) \, (\partial _{\mu}U \partial ^{\mu} U^{\dagger}+r(MU^{\dagger}+UM))^{dq} \, U^{qs}
\ + \ O(p^{6})
\label{eq:LDevol}
\end{eqnarray}
with $a = -3 \Lambda^{2} / (4 \pi F_{\pi})^{2}$,
 $b = \ln (\Lambda^{2}/m_{\pi}^{2}) / 32 \pi^{2}$ and $\Lambda$ is the ultra-violet cut-off.

We see that the lowest order non-linear realization of the TA operator, obtained by factorizing the quark from the gluon fields in (\ref{eq:QGTA}), does indeed
show up in the second line of Eq. (\ref{eq:LDevol}).

\section{Conclusion}

We have indicated how a potentially important contribution of the QCD trace anomaly
to the $\Delta I=1/2$ Kaon decay amplitudes can arise from strong interaction corrections to the
$\Delta S =1$ Fermi Hamiltonian.
A more quantitative analysis will require the investigation of the matching between the SD and LD evolutions.
A previous study\,\cite{Penin94} of $K \ra \pi \pi$ decay amplitudes with gluons in the intermediate state
(but without any reference to the trace anomaly) predicts an enhancement of $14\pm 6\%$ using QCD sum rules. The consistency between the two methods would be worth checking.

\section*{Acknowledgments}

This work has been done in collaboration with J.-M. G\'erard.\,\cite{Gerard03} Financial supports from IISN
and Interuniversity Attraction Pole P5/27 are also acknowledged.

\section*{References}

\end{document}